# A $\delta$-adjusted Method for Tipping Point Sensitivity Analysis for a Binary Outcome with Missing Values

**Mallikarjuna Rettiganti, Yipeng Wang, and Yongming Qu**

Global Statistical Sciences
Eli Lilly and Company, Indianapolis, IN, USA

Correspondence to: Yongming Qu, Lilly Corporate Center, Indianapolis, IN 46285.
Email: qu_yongming@lilly.com

## Abstract

Missing data remain a persistent challenge in clinical trials despite preventive efforts. Handling missing data in clinical trials requires not only a primary analysis but also sensitivity analyses, as the underlying missingness assumptions are difficult to verify. The δ-adjusted analysis, which assesses the robustness of primary results by adding a sensitivity parameter δ for one or both treatment groups, has been widely used for continuous outcomes. Then, a set of tipping points which flip the significance of the primary analysis can be identified. However, there is no corresponding $\delta$-adjusted method for binary variables. In this article, we propose a δ-adjusted method for binary variables that addresses this gap. This method assesses robustness by adding a sensitivity parameter to the logit function of the estimated response probability under the primary imputation approach and shares the same property as the δ-adjusted method for continuous variables. We applied this method to a clinical trial evaluating the treatment effect of tirzepatide 15 mg versus 5 mg on a composite binary endpoint. This method provides a principled sensitivity analysis framework for binary endpoints with broad applicability in clinical trials.

**Keywords:** Missing data, sensitivity analysis, tipping point, binary outcome, multiple imputation, clinical trials

# 1. Introduction

Missing data are ubiquitous in clinical trials despite rigorous protocols designed to minimize their occurrence (National Research Council, 2010). The reasons for missing values are diverse and can be categorized into several types: (1) participants remain in the study but their data are not collected due to operational issues, such as compromised sample storage or equipment malfunction; (2) participants withdraw from the study, leading to missing values for all subsequent planned visits and measurements; (3) data are censored when a hypothetical strategy is employed to handle certain intercurrent events as defined in the ICH E9(R1) addendum (ICH E9(R1), 2019). For example, data collected after the use of rescue medication may be treated as missing under certain estimands. These latter missing values are often referred to as unobservable outcomes in the estimand framework.

The appropriate handling of missing data is critical for valid inference in clinical trials. The choice of methods or models for addressing missing values should align with the estimand of interest and the potential outcomes framework (Qu and Lipkovich, 2021). For missing values not attributable to intercurrent events, when the potential outcome represents the value had participants adhered to the randomized treatment, and when missing values can be adequately modeled using observed data, a missing-at-random (MAR) assumption may be invoked for imputation. Conversely, if the potential outcome corresponds to a treatment policy strategy (typically reflecting the outcome in the presence of intercurrent events), a pattern mixture model under the assumption of missingness not at random (MNAR) may be more appropriate (Little, 1993; Hedeker and Gibbons, 1997).

Regardless of the primary approach used to handle missing data, sensitivity analyses are invariably recommended to demonstrate the robustness of primary analysis results (National Research Council, 2010; ICH E9(R1), 2019; European Medicines Agency, 2010). The National Research Council (2010) emphasized that "the analysis under the MAR assumption should be the primary analysis but should be supplemented by sensitivity analyses under MNAR assumptions." Sensitivity analyses serve to explore how conclusions might change under alternative, often more pessimistic, assumptions about the missing data mechanism.

Among the various sensitivity analysis approaches, δ-adjusted analysis and tipping point analysis have gained widespread acceptance (Yan et al., 2009; Ratitch et al., 2013; Carpenter et al., 2013). For continuous outcomes where smaller values indicate better outcomes, the delta-adjusted method adds a sensitivity parameter $\delta > 0$ to imputed values in a treatment group after the primary imputation. The analysis is then repeated using the same statistical model as the primary analysis. By systematically varying $\delta$, investigators can identify the value at which the conclusion "tips" from being statistically significant to being insignificant, the so-called tipping point. If the tipping point corresponds to a clinically implausible degree of departure from the MAR assumption, the primary analysis results can be considered robust. This approach has been extended to two-way sensitivity analyses, where distinct shift parameters $\delta_1 < 0$ and $\delta_2 > 0$ are applied to the control and experimental groups, respectively, resulting in tipping regions that typically form lines or elliptical segments in the $(\delta_1, \delta_2)$ parameter space (Ratitch et al., 2013).

The implementation of tipping point analysis for continuous variables is conceptually straightforward and has been further refined in recent years. Gorst-Rasmussen and Tarp-Johansen (2022) proposed a computationally efficient fast tipping point analysis that analytically expresses the mean and variance of the δ-adjusted treatment effect as a function of $\delta$ and the treatment effect estimate without delta-adjustment. This eliminates the need for repeated imputations across the $\delta$ grid. Torres et al. (2025) introduced an imputation-free tipping point approach that requires minimal assumptions, basing inference on observed data and assumed mean differences between dropouts and completers within each treatment arm. Cro et al. (2020) provided practical guidance on implementing reference-based imputation methods, which represent a special class of MNAR assumptions where missing data in one treatment group are imputed using parameters from a reference group. García-Hernandez et al (2023) and Luo et al (2025) provided a framework for implementing some pattern-mixture model and tipping-point analysis analytically without requiring imputation.

For binary outcomes, the landscape is more complex and relatively less well-developed. Several approaches have been proposed, but each suffers from important limitations. Yan et al. (2009) first proposed to summarize tipping points as number of successes (or failures) on subjects with missing data in each arm that would change the study's conclusion. However, their method does not account for imputation uncertainty and cannot adjust for covariates, limiting its practical utility.

Liublinska and Rubin (2014) proposed an enhanced δ-adjusted method by adding a shift parameter (δ) on the logit probability of success (estimated under MAR assumption) for each treatment group, mimicking the δ-adjusted method for the continuous variables. This method accounts for the imputation variability in drawing inferences with added shift parameters. However, the tipping point is still displayed by in numbers of additional counts of successes in each arm as used in Yan et al (2009), instead of the shift parameters themselves. Thus, the tipping points themselves are based on the counts of successes, while the shift parameters assumed under the MNAR assumptions only appear through ranges of counts superimposed on the display through rectangles which are then compared with tipping points to assess robustness. This makes it harder to assess the exact tipping points in terms of the shift parameters that might reverse the conclusions of the study.

Sui et al. (2023) proposed generating binary outcomes from Bernoulli distributions with probabilities varied across a grid of sensitivity parameters to overcome the limitation of a constant shift parameter applied on the logit scale not translating into a constant shift in probability due to the non-linearity of the logit function. Their approach assumes a constant response probability across all subjects with missing data, independent of subject-specific covariates - an assumption that may be unrealistic in many settings. More importantly, this method suffers from a fundamental limitation: the delta-adjustment is decoupled from the primary imputation model. Unlike continuous outcome settings where $\delta = 0$ recovers the primary analysis exactly, these methods for binary outcomes do not maintain this coherence property, potentially undermining the conceptual and interpretive connection between the primary and sensitivity analyses.

When a binary variable is derived from a single underlying continuous variable (e.g., response defined as achieving a certain threshold on a continuous scale), the recommended approach is to impute the continuous variable and then derive the binary outcome (Ma et al., 2022). In such cases, δ-adjusted analysis can be performed and summarized at the continuous variable level. However, this strategy is not applicable when the binary outcome is directly collected (e.g., clinical remission: yes/no) or when it is derived from multiple continuous variables using complex criteria (e.g., composite endpoints requiring improvements across multiple domains). When binary outcomes are derived from multiple underlying continuous variables (e.g., composite endpoints requiring improvements across multiple domains), conducting the δ-adjusted analysis at the level of the individual continuous variables is not feasible. Such an approach would require specifying a separate shift parameter for each continuous variable, resulting in a multi-dimensional tipping point surface rather than a simple tipping point boundary. The dimensionality of this surface grows with the number of variables used to define the binary outcome, rendering the results increasingly difficult to summarize, visualize, and interpret.

In this article, we propose a general delta-adjusted tipping point analysis for binary outcomes that is directly analogous to methods for continuous variables and shares similar desirable properties, particularly the coherence property where $\delta = 0$ recovers the primary analysis. Our approach applies a uniform shift parameter $\delta$ on the logit scale, providing a natural extension of the additive shift for continuous outcomes. The main contributions of this article are threefold. First, tipping points are identified and displayed directly in the space of the sensitivity parameters $(\delta_1, \delta_2)$, with multiple imputation inference combined by Rubin's rules at each point and with the primary analysis located at the origin, in contrast to displays based on counts of successes among participants with missing data (Yan et al., 2009; Liublinska and Rubin, 2014). Second, the method applies both to directly collected binary endpoints and to binary endpoints derived from one or more continuous variables, irrespective of the model used to impute the underlying continuous variables; for composite endpoints derived from multiple continuous variables, δ-adjustments applied at the level of the individual continuous variables do not yield a simple, interpretable tipping boundary. Third, the method preserves the coherence property that $\delta = 0$ recovers the primary analysis, which is not shared by approaches that decouple the δ-adjustment from the primary imputation model (Sui et al., 2023). In Section 2, we describe the methodology for both directly collected binary variables and binary variables derived from one or more continuous measures. In Section 3, we illustrate the method using data from the SURMOUNT-1 trial, a phase 3 study evaluating an obesity management medication. Finally, in Section 4, we provide a summary and discussion of the method's advantages, limitations, and practical considerations.

## 2. Methods

Binary variables in clinical trials can be classified into two primary types. First, a binary variable may be directly collected during the trial. For example, in a fixed-duration study, the primary endpoint might assess whether a patient's condition has achieved a predefined improvement threshold (yes/no) at a specific timepoint, such as headache freedom at 2 hours in migraine trials or clinical remission at 12 weeks in inflammatory bowel disease studies. Second, binary variables may be derived from one or more continuous variables. This approach is common in immunology and rheumatology, exemplified by PASI-75 (75% improvement in Psoriasis Area and Severity Index) for psoriasis and ACR20 (20% reduction in tender and swollen joint counts plus 20% improvement in at least three of five key domains: pain, function, patient/physician global assessment, and inflammatory markers) for rheumatoid arthritis. We describe tipping point analysis methods tailored to each type of binary variable.

## 2.1. Directly Collected Binary Variables

Let $Y_{ij}$ denote the binary outcome for subject $j$ in treatment group $i$. Without loss of generality, we consider two treatment groups, where $i = 1$ represents the control group and $i = 2$ represents the experimental treatment group. Let $X_{ij}$ denote the vector of baseline covariates and $Z_{ij}$ denote the vector of ancillary variables used in the imputation model. The "complete" data after imputation can be expressed as

$$Y_{ij}^{C,m} = R_{ij}Y_{ij}^{o} + \left(1 - R_{ij}\right)Y_{ij}^{m}, \quad m = 1,2 \dots, M,$$

where $Y_{ij}^{o}$ is the observed value for the outcome, $R_{ij}$ is the missingness indicator (1 for observed and 0 for missing), and $Y_{ij}^{m}$ is the imputed value at the $m$-th imputation.

The multiple imputation in this case is typically based on an underlying logistic regression model:

$$\Pr(Y_{ij} = 1|X_{ij}, Z_{ij}^{o}) = g^{-1}\left(\gamma_0 + \gamma_{x,i}'X_{ij} + \gamma_{z,i}'Z_{ij}^{o}\right), \quad (1)$$

where $g^{-1}(x) = \exp(x)/\left(1 + \exp(x)\right)$ is the inverse logit function. The primary analysis will proceed by analyzing the data on each of the multiply imputed datasets using an appropriate analysis method and combining the results across all $M$ imputations according to Rubin’s rules.

The two-way δ-adjusted sensitivity analysis with $M$ imputations is then conducted as follows:

For $m = 1,2, \dots, M$, generate parameter draws from the distribution:

$$\left(\tilde{\gamma}_0^{m}, \tilde{\gamma}_{x,i}^{m\prime}, \tilde{\gamma}_{z,i}^{m\prime}\right)' \sim N\left(\left(\hat{\gamma}_0, \hat{\gamma}_{x,i}', \hat{\gamma}_{z,i}'\right)', \hat{V}\left[\left(\hat{\gamma}_0, \hat{\gamma}_{x,i}', \hat{\gamma}_{z,i}'\right)'\right]\right), \quad (2)$$

where $\left(\hat{\gamma}_0, \hat{\gamma}_{x,i}', \hat{\gamma}_{z,i}'\right)'$ is the estimate of the parameters $\left(\gamma_0, \gamma_{x,i}', \gamma_{z,i}'\right)'$ and $\hat{V}\left[\left(\hat{\gamma}_0, \hat{\gamma}_{x,i}', \hat{\gamma}_{z,i}'\right)'\right]$ is the corresponding variance estimate which are obtained from the multiple imputation model on participants with observed outcomes.

2. For each imputation $m$, and each participant with missing data, calculate the δ-adjusted response probability:

$$\tilde{p}_{ij}^{m} = g^{-1}\left(\tilde{\gamma}_0 + \tilde{\gamma}_{x,i}'X_{ij} + \tilde{\gamma}_{z,i}'Z_{ij}^{o} + \delta_i\right).$$

3. For each imputation $m$, generate a random variable $Y_{ij}^m$ from a Bernoulli distribution with mean $\tilde{p}_{ij}^m$.
4. Analyze each of the M imputed datasets using the same model as the primary analysis (e.g., logistic regression).
5. Combine results across all M imputations within each combination of δ values using Rubin's rules (Rubin, 1987).

Note that using $\delta_i = 0, i = 1, 2$ in Step 2 recovers the primary analysis which is a desired coherence property. In this setting, the property holds exactly because, with $\delta_1 = \delta_2 = 0$, Steps 1–5 reduce to the primary multiple imputation analysis.

Although we present the method using the logit link, the procedure does not depend on this choice. More generally, $g$ in Step 2 may be any strictly monotone link function mapping (0, 1) onto the real line, such as the probit or complementary log-log link; $\delta_i$ is then added to the response probability predicted by the primary imputation model after transformation by $g$, and the result is back-transformed using $g^{-1}$. The coherence property holds for any such choice of $g$. We recommend the logit link as the default because $\exp(\delta_i)$ then has a direct interpretation as the multiplicative change in the odds of response among participants with missing data in treatment group $i$, relative to the odds predicted by the primary imputation model, which is on the same scale as the odds ratio typically reported in the primary analysis. Under the probit link, for example, $\delta_i$ is a shift on the latent standard normal scale, which lacks a comparably direct clinical interpretation. Furthermore, the δ-adjustment does not require the imputation model itself to be a logistic regression model; the shift serves only to parameterize the departure from the primary imputation model. This is illustrated in Section 2.2, where the underlying continuous variables may be imputed using any appropriate model (e.g., the multivariate normal model used in Section 3).

## 2.2. Binary Variables Derived from One or More Continuous Variables

In cases where $Y_{ij}$ is derived from a vector of continuous variables $C_{ij}$:

$$Y_{ij} = h(C_{ij}; \tau),$$

where $\tau$ is a given vector of threshold values that define the binary response from the set of continuous variables. In the primary analysis, the continuous variables are imputed using standard software packages such as PROC MI in SAS® (SAS Institute Inc., 2024) or the *mice* package in R (van Buuren and Groothuis-Oudshoorn, 2011) as $C_{ij}^m$ $(m = 1, 2, \ldots, M)$. The corresponding imputed value for the binary response variable is then

$$Y_{ij}^m = h(C_{ij}^m; \tau).$$

For the primary analysis, data from each imputed dataset is analyzed according to an appropriate method such as logistic regression analysis and the results across all imputations are combined using Rubin's rules.

Following the primary multiple imputation analysis, the two-way δ-adjusted imputation for sensitivity analysis is then conducted as follows:

1. Estimate the probability that subject $j$ in treatment group $i$ has a response of 1 as

$$\tilde{p}_{ij} = M^{-1} \sum_{m=1}^{M} Y_{ij}^{m}.$$

2. Calculate the δ-adjusted probability for subject $j$ in treatment group $i$ as

$$\tilde{p}_{ij}(\delta_i) = g^{-1}\left(\log\left(\frac{\tilde{p}_{ij}}{1-\tilde{p}_{ij}}\right) + \delta_i\right),$$

   where $g^{-1}(x) = \exp(x)/\big(1 + \exp(x)\big)$.

3. Generate $M$ imputed responses $Y_{ij}^{m}(\delta_i), m = 1, 2, \ldots, M,$ from a Bernoulli distribution with mean $\tilde{p}_{ij}(\delta_i)$.

4. For each combination of $\delta$ values applied to the two treatment arms, and for each of the $M$ imputations, analyze the data using logistic regression or an appropriate modeling approach to compare outcomes between treatment groups.

5. As in Section 2.1, combine results across all $M$ imputations within each combination of $\delta$ values using Rubin's rules.

Note that the proposed approach does not combine the underlying continuous variables into a single continuous variable. The binary response is first derived from the imputed continuous variables through $h(\cdot;\tau)$, exactly as in the primary analysis, and the sensitivity parameter $\delta_i$ is then applied to the logit of each participant's estimated response probability $\tilde{p}_{ij}$. The departure from the primary imputation assumption is therefore expressed directly on the scale of the binary endpoint that defines the estimand, regardless of the number of underlying continuous variables or the complexity of $h(\cdot;\tau)$. With $\delta_1 = \delta_2 = 0$, each δ-adjusted imputed response is drawn with probability $\tilde{p}_{ij}$, the average of the participant's primary imputed responses, so the primary analysis is recovered approximately, with differences arising from Monte Carlo error and from the approximation described below (see Section 3.3 for a numerical illustration).

It is technically possible but computationally intensive to estimate the variability and covariance in $E\big[Y_{ij}^{m}\big]$ analogous to Equation (2) for directly collected binary variables. In practice, we ignore this source of variability in the δ-adjusted imputation. Because the variability in $E\big[Y_{ij}^{m}\big]$ is of order $n^{-1}$, the majority of the variability in $Y_{ij}^{m}$ arises $V\big(Y_{ij}^{m}\big|E\big[Y_{ij}^{m}\big]\big)$. Ignoring the between-imputation variability in $E\big[Y_{ij}^{m}\big]$ has only a minimal effect on the imputation results and does not substantially impact inference.

The results from a two-way $\delta$-adjusted tipping point sensitivity analysis can be effectively summarized using a graphical grid display, commonly presented as a heatmap. In this display, values of $\delta_1$ are arranged along the x-axis and values of $\delta_2$ along the y-axis, with each cell in the grid representing a specific combination of $(\delta_1, \delta_2)$. For each combination, the corresponding p-value and treatment effect estimate (e.g., odds ratio) are computed following the procedures in Sections 2.1 or 2.2, and these values are displayed within each cell.

## 3. Real Data Application

The SURMOUNT-1 study (Jastreboff et al., 2022) was a phase 3, double-blind, randomized controlled trial that randomized 2,539 adults with obesity (BMI ≥30 $kg/m^2$) or overweight (BMI ≥27 $kg/m^2$) with at least one weight-related comorbidity to receive either placebo or once-weekly subcutaneous tirzepatide at doses of 5 mg, 10 mg, or 15 mg. The co-primary endpoints were percent change from baseline in body weight and a categorical response of ≥5% weight reduction after 72 weeks of treatment. We apply the method proposed in Section 2 to data from the SURMOUNT-1 study. The study was conducted in accordance with the Declaration of Helsinki guidelines on good clinical practices (World Medical Association 2010).

### 3.1. A Composite Binary Variable Derived from Multiple Continuous Variables

In this example, the composite binary response is based on three underlying continuous outcomes: percent change from baseline to Week 72 in body weight, change from baseline in systolic blood pressure (mmHg), and change from baseline in diastolic blood pressure (mmHg). A subject is classified as a responder to the composite binary outcome (i.e., the value is 1) if the subject achieves at least a 15% reduction in body weight from baseline *and* achieves normalization or clinical improvement in blood pressure. Normalization is defined as achieving systolic and diastolic blood pressure values below 130 mmHg and 80 mmHg, respectively, at Week 72. Clinical improvement is defined as at least a 7-mmHg improvement from baseline in systolic blood pressure or a 5-mmHg improvement in diastolic blood pressure, with no clinical worsening in the other component (diastolic or systolic, respectively).

We restricted the analysis to subjects from the SURMOUNT-1 study with baseline systolic blood pressure ≥130 mmHg or diastolic blood pressure ≥ 80 mmHg. To demonstrate practical tipping points for this sensitivity analysis, the primary comparison focuses on the derived composite outcome between the 5 mg and 15 mg tirzepatide arms. Specifically, 349 patients randomized to tirzepatide 5 mg and 350 patients randomized to tirzepatide 15 mg were included in this analysis. A total of 40 participants (11.5%) in the 5 mg arm and 28 participants (8.0%) in the 15 mg arm had missing outcomes for the composite endpoint. Among the 309 participants with complete data in the 5 mg arm, 143 (46.3%) had a positive response to the composite, while among the 322 participants with complete data in the 15 mg arm, 189 (58.7%) had a positive response.

## 3.2. Primary Analysis Based on Multiple Imputation

The primary analysis was aligned with the estimand of interest; for illustration, the analysis of the composite binary outcome was conducted under the efficacy estimand. All missing data were handled using the hypothetical strategy, which assumes missingness is independent of the outcome conditional on the observed covariates (i.e., MAR assumption). The primary analysis for the composite binary outcome proceeded as follows:

i) For participants with missing outcomes, the three underlying continuous variables were imputed 100 times using a Markov Chain Monte Carlo procedure under the assumption of a multivariate normal distribution. The imputation model included age, baseline BMI, body weight at baseline and Week 72, systolic and diastolic blood pressure at baseline and Week 72, triglyceride levels at baseline and Week 72, baseline heart rate, and baseline HbA1c.

ii) For each imputed dataset, the composite response variable was derived, and the treatment comparison between the 15 mg and 5 mg arms was evaluated using logistic regression. The model specified the composite binary outcome as the dependent variable, the treatment indicator (15 mg vs. 5 mg) as the primary explanatory variable, and included baseline age, gender, body weight, triglycerides, systolic blood pressure, and diastolic blood pressure as covariates.

iii) The results from the logistic regression for each of the 100 imputations were combined using Rubin's rules (Rubin, 1987).

The pooled treatment effect estimate, obtained by combining 100 imputed datasets via Rubin's rules, indicated that participants in the 15 mg arm had significantly higher odds of achieving the composite binary outcome compared to the 5 mg arm (OR = 1.82, 95% CI: 1.31–2.54, p = 0.0003).

## 3.3. Tipping Point Sensitivity Analysis

To assess the robustness of the primary analysis results to potential departures from the MAR assumption, we conducted a two-way δ-adjusted tipping point analysis based on the method described in Section 2.2. For each subject with a missing composite outcome, we added sensitivity parameters $\delta_1$ and $\delta_2$ to the logit-transformed predicted probabilities in the 5 mg and 15 mg treatment arms, respectively. In this case, $\delta_1$ was positive and $\delta_2$ was negative, thus narrowing the difference between the two treatment groups and representing a conservative sensitivity analysis. Specifically, $\delta_1$ was increased from 0 to 3 in increments of 0.2, and $\delta_2$ was decreased from 0 to -3 in increments of -0.2.

For each combination of $(\delta_{1,}\delta_2)$, we generated 100 δ-adjusted imputations by drawing binary outcomes from Bernoulli distributions with the adjusted predicted probabilities. Within each imputation, logistic regression analysis was performed using the same model specification as the primary analysis, adjusting for baseline age, gender, body weight, triglycerides, systolic blood pressure, and diastolic blood pressure. Results across the 100 imputations were then combined using Rubin's rules to obtain pooled treatment effect estimates, standard errors, and p-values for each $\delta$ combination.

Figure 1 presents the results of the tipping point analysis as a grid plot, with the upper values in each cell showing p-values and lower values showing odds ratios for the treatment comparison (15 mg vs. 5 mg). Under the primary MAR assumption ($\delta_1 = \delta_2 = 0$), the treatment effect remained highly significant (*p* = 0.0003, OR = 1.83), consistent with the primary analysis results (OR = 1.82, *p* = 0.0003). Cyan-shaded cells indicate $\delta$ combinations where the treatment effect was statistically significant ($p < 0.05$), while red-shaded cells represent combinations where significance was lost ($p \geq 0.05$). The analysis illustrates substantial robustness of the primary findings, with statistical significance maintained across most of the sensitivity parameter space. The tipping point region, where the treatment effect ceased to be significant, began to emerge at combinations of large sensitivity parameters, approximately when $\delta_1 \geq 1.6$ and $\delta_2 \leq -1.4$ simultaneously. This suggests that a change in the statistical significance of the treatment effect would occur only under parameter shifts in which the odds of an imputed response in the 5 mg arm are increased by a factor of $\exp(1.6) \approx 4.95$ and the odds of an imputed response in the 15 mg arm are reduced by a factor of $\exp(-1.4) \approx 0.25$ (equivalently, divided by 4.05) than those expected by imputation under the MAR. These large departures support the robustness of the primary analysis conclusions.

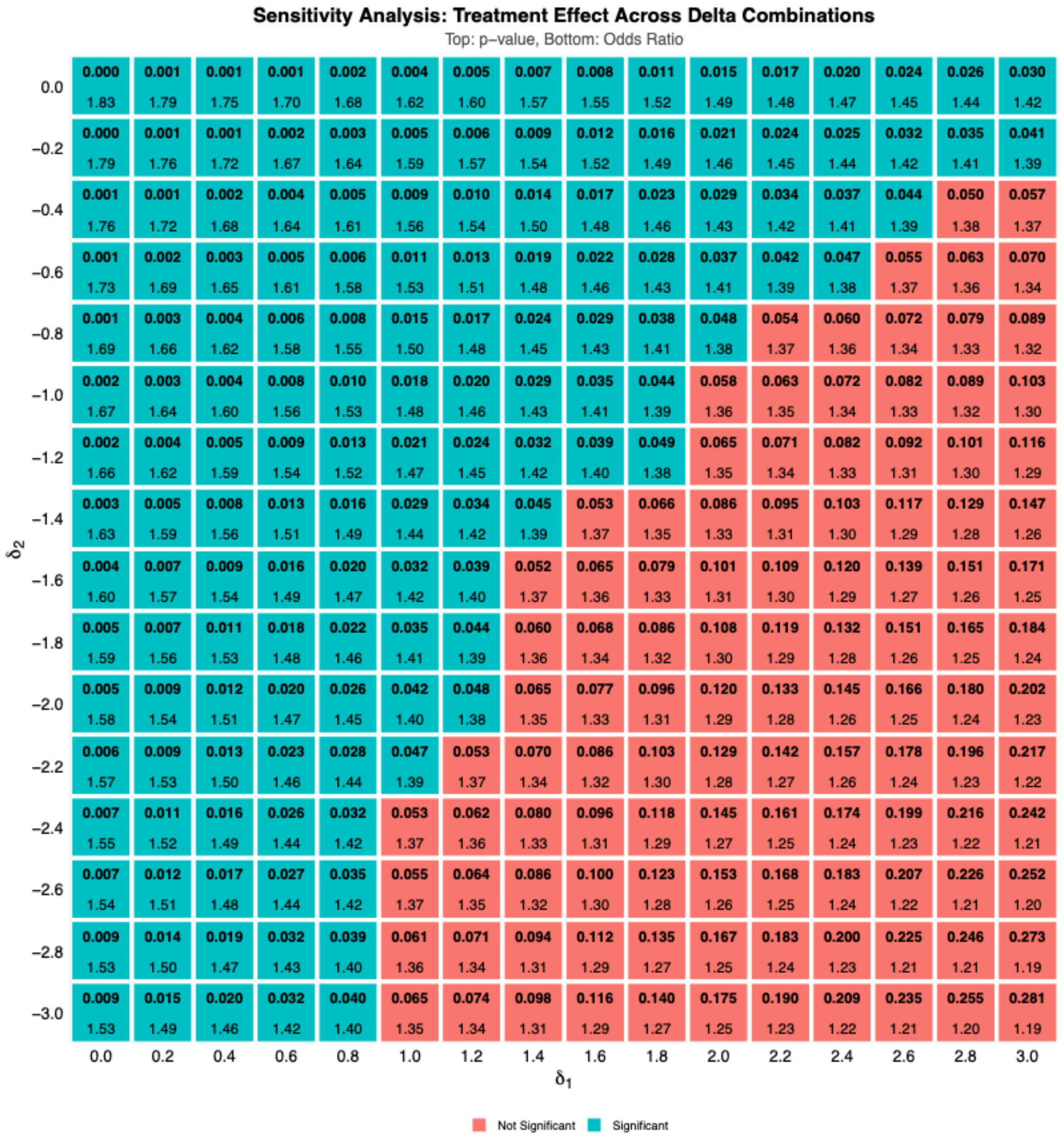

**Figure 1.** Two-way tipping point sensitivity analysis for the treatment effect (15 mg vs. 5 mg). The grid plot presents p-values (upper value in each cell) and odds ratios (lower value in each cell) across sensitivity parameter combinations $(\delta_{1,}\ \delta_2)$. Cyan shading indicates statistically significant treatment effects ($p < 0.05$); red shading indicates non-significant effects ($p \geq 0.05$). The baseline MAR assumption corresponds to $\delta_1 = \delta_2 = 0$. For each $\delta$ combination, results were pooled from 100 imputations using Rubin's rules.

As a comparison, the ETP display method proposed by Liublinska and Rubin (2014) was applied to the data from the SURMOUNT-1 study introduced in Section 3.1 and surrounding discussion was included in the Appendix.

## 4. Summary and Discussion

In this article, we proposed a general δ-adjusted tipping point sensitivity analysis method for binary outcomes in clinical trials with missing data. The method offers several important advantages. First, by operating on the logit scale, the delta-adjustment provides a natural and interpretable parameterization of departures from the primary analysis. The logit transformation ensures that adjusted probabilities remain bounded between 0 and 1, and the multiplicative effect on odds ($\exp(\delta)$) has a straightforward clinical interpretation. While Liublinska and Rubin (2014) incorporate a similar method for directly collected binary outcomes, their enhanced tipping point display does not directly incorporate the sensitivity parameters as part of the display thus making it hard to understand the magnitude of deviation from the primary imputation method that is necessary to tip the results to being statistically insignificant. In addition, because each cell of the ETP grid is defined only by the numbers of successes among participants with missing data, significance in each cell is evaluated from a single completed 2 × 2 table using an unadjusted test; the resulting tipping boundary therefore need not correspond to that of a covariate-adjusted, multiple imputation-based primary analysis, and the display combines imputation variability with departure from the MAR assumption (see the Appendix). Second, the method seamlessly accommodates both one-way and two-way tipping point analyses, enabling investigators to explore sensitivity to MNAR assumptions in one or both treatment arms. Third, the approach is flexible enough to handle both directly collected binary outcomes and binary outcomes derived from continuous variables, making it widely applicable across diverse clinical trial settings. Fourth, by incorporating the delta-adjustment within the multiple imputation framework and combining results using Rubin's rules, the method properly accounts for imputation uncertainty. The real data application from the SURMOUNT-1 trial demonstrates the practical utility of the proposed method. In this application, the primary analysis is based on imputing data under the MAR assumption to estimate the estimand with all intercurrent events handled using a hypothetical strategy. Recently, several publications advocated the treatment policy estimand, an estimand with intercurrent events handled by the treatment policy strategy (Fleming, et al 2025; Troxel, et al 2026). Our proposed method can also be similarly applied to a different imputation method that is consistent with the treatment policy estimand without modification.

For binary variables derived from continuous variables (Section 2.2), we ignore the variability in the estimated response probabilities $\tilde{p}_{ij}$ when applying the delta-adjustment. This simplification is justified because the variability in $\tilde{p}_{ij}$ is of order $n^{-1}$, and the dominant source of variability comes from the Bernoulli sampling variability conditional on the mean probability. In the application, the *p*-value using the proposed δ-adjusted method when $\delta_1 = \delta_2 = 0$ was 0.0003 (OR = 1.83), virtually identical to the primary analysis *p*-value of 0.0003 (OR = 1.82). This indicates that such a simplification is unlikely to severely underestimate the variance for the δ-adjusted estimators.

For binary endpoints derived from multiple continuous variables, a single sensitivity parameter per treatment arm expresses the departure from the primary imputation assumption on the scale of the composite endpoint. Specifying separate shift parameters for each underlying continuous variable would instead lead to a multi-dimensional tipping surface, with shift parameters on different scales that are difficult to compare and interpret. A limitation of the proposed approach is that a single $\delta$ per arm does not attribute the departure from the primary assumption to any one component. When component-specific departures are of clinical interest, component-level δ-adjustments can be performed as supplementary analyses.

In conclusion, the δ-adjusted sensitivity analysis method for binary outcomes proposed in this article provides a principled and practical framework for sensitivity analysis in clinical trials. It represents an important addition to the statistical toolkit for handling missing binary outcomes, complementing existing approaches and filling a methodological gap in the literature.

## Data Availability

The clinical trial data used in this manuscript are proprietary to Eli Lilly and Company and cannot be made publicly available due to patient confidentiality obligations and data use agreements governing their use.

## Ethical Statement

The real-data application uses de-identified data from SURMOUNT-1 (NCT04184622), a trial conducted in accordance with the Declaration of Helsinki and Good Clinical Practice guidelines with ethics committee/IRB approval at each site and written informed consent from all participants. This secondary methodological analysis did not require additional ethical approval.

## Conflict of Interest

Mallikarjuna Rettiganti, Yipeng Wang, and Yongming Qu are employees and minority shareholders of Eli Lilly and Company.

## Funding

The authors received no specific funding for this work beyond their employment by Eli Lilly and Company.

## Author Contributions

Mallikarjuna Rettiganti: Conceptualization, methodology, program, writing original draft, review and editing.

Yipeng Wang: Data analysis/program, writing original draft, review and editing.

Yongming Qu: Conceptualization, methodology, writing original draft, review and editing, supervision.

## Acknowledgments

The authors thank Jit Mitra for his thoughtful review and comments that helped improve this manuscript.

## Appendix

**Real Data Application – Enhanced Tipping Point display**

As a way of comparing the method for tipping point proposed in this manuscript with the enhanced tipping point display (ETP) method described in Liublinska and Rubin (2014), we applied the ETP display method (Figure A1) on the SURMOUNT-1 study data used in Section 3. This method uses the deterministic grid of successes in each arm (5 mg arm on the x-axis and 15 mg arm on the y-axis) that are needed to overturn the significance of the primary analysis. The blue shaded region shows the region of statistical significance, and the green shaded region shows the region of insignificance. The boundary separating the two regions is shown using a red line. The region enclosed by the blue box indicates the combinations of successes in the two arms that can result when missing data are imputed under the missing at random assumption (MAR), which mostly lies in the region of statistical significance. The region enclosed by the orange box shows the combination of counts in each arm when missing data are imputed under the missing not at random assumption (MNAR) where $\delta_1$=1.6 and $\delta_2$= -1.4, where $\delta$'s are the sensitivity parameters as proposed in this manuscript. Following Liublinska and Rubin (2014), each box spans the minimum to the maximum numbers of successes among participants with missing data in each arm across the 100 imputations under the corresponding assumption. Figure 1 suggests this combination of $\delta$'s is on the boundary where the treatment effect becoming statistically insignificant. For the same combination of $\delta$'s the ETP display in Figure A1 shows the range of counts of successes in each arm obtained across the imputations (orange box) which suggests this combination of $\delta$'s is most likely to tip the results to becoming statistically insignificant. The purple and orange dashed lines represent the expected number of successes in each arm under the MAR and MNAR assumptions respectively.

The ETP display has the appealing feature of presenting sensitivity results in terms of the raw counts of successes among participants with missing data, which is transparent for small trials with unadjusted analyses. However, several features limit its use as a tipping point analysis for the primary analysis of a confirmatory trial. First, each cell of the ETP grid treats a single completed set of counts as fully observed and evaluates significance with an unadjusted two-sample test (here, a two-sided z-test). The boundary in Figure A1 is therefore the tipping boundary of a simpler analysis rather than that of the prespecified covariate-adjusted logistic regression combined across imputations using Rubin's rules; when covariates are included in the analysis model, the number of imputed successes does not by itself determine the result. Consequently, the primary analysis result does not correspond to any point on the display. Second, because the boxes represent the spread of the imputed counts, the ETP combines imputation variability with departure from the MAR assumption. In Figure A1, a small portion of the MAR box lies in the region of insignificance, although the primary analysis gives *p* = 0.0003, so the display can suggest fragility even under the primary assumption. Third, when a box straddles the boundary, as the MNAR box for $\delta_1 = 1.6$ and $\delta_2 = -1.4$ does, the display offers no single conclusion for that assumption, whereas the proposed method yields one pooled inference for each $(\delta_1, \delta_2)$. Fourth, because the axes are counts, the display depends on the number of participants with missing data, so the same departure from MAR appears differently across trials or endpoints, and the grid becomes dense as the amount of missing data increases. Finally, the ETP evaluates one assumption at a time, each requiring its own overlaid box, which can make the display crowded when several MNAR assumptions are considered, whereas the proposed display identifies the entire tipping region in $(\delta_1, \delta_2)$ directly, with the primary analysis at the origin. We did not separately evaluate the approach of Yan et al. (2009), because its display corresponds to the ETP grid in Figure A1 without the boxes representing imputation-based assumptions.

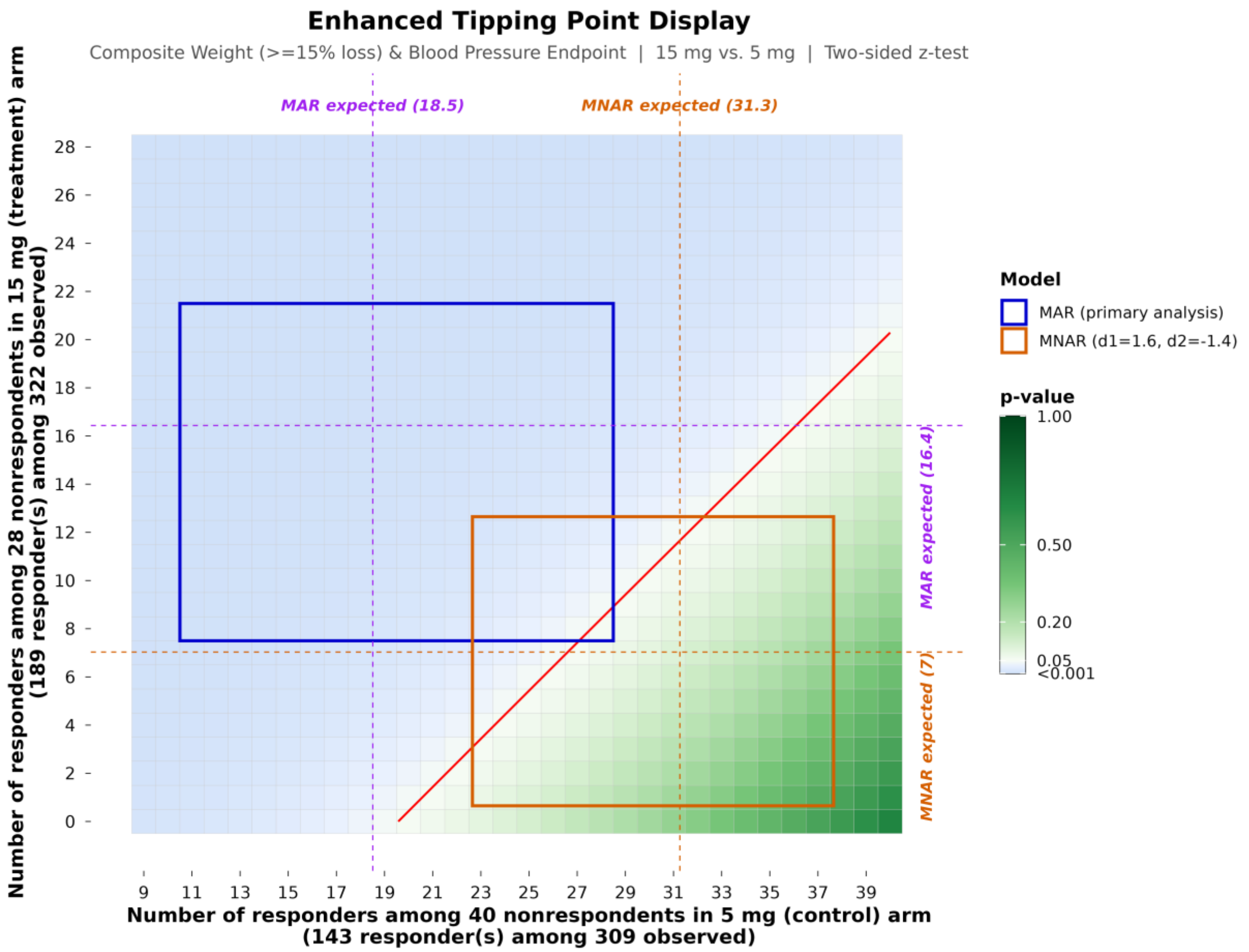

*Figure A1. ETP display method from Liublinska and Rubin applied to data from the SURMOUNT-1 trial of tirzepatide comparing the 15mg arm vs. the 5 mg arm. The blue shaded region indicates region of statistical significance while the green shaded region indicates region of statistical insignificance. The blue box indicates the combination of successes in each arm assuming MAR imputation while the orange box indicates possible counts under the MNAR imputation where* $\delta_1 = 1.6$ *and* $\delta_2 = -1.4$*. Here, the* $\delta$*'s are the sensitivity parameters applied on the logit of the probabilities in each arm before imputation as described in Section 2. The purple and orange dashed lines indicate the expected numbers of successes among participants with missing data under the MAR and MNAR assumptions, respectively. Each box spans the minimum to the maximum numbers of successes across the 100 imputations under the corresponding assumption. Significance in each cell of the grid is based on an unadjusted two-sided z-test applied to the completed counts, whereas Figure 1 is based on the covariate-adjusted logistic regression combined across imputations using Rubin's rules.*